\documentclass[twocolumn,epjc3,final]{svjour3}
\RequirePackage[T1]{fontenc}

\smartqed  

\RequirePackage{graphicx}
\RequirePackage{mathptmx}      
\RequirePackage{flushend}
\RequirePackage[numbers,sort&compress]{natbib}

\usepackage{amsmath, txfonts,multirow}
\usepackage{graphicx,bm,booktabs}
\usepackage{color}
\usepackage[bookmarks=true,bookmarksopen=false,plainpages=false,breaklinks=true,
   bookmarksnumbered=true,hypertexnames=false,
   filecolor=blue,urlcolor=three,menucolor=three,
   linkcolor=three,citecolor=blueone, colorlinks,
   anchorcolor=blue,runcolor=pink,frenchlinks=red
   pdfstartview=FitH,pdftitle=title,%
   pdfauthor=author]{hyperref}
   
\allowdisplaybreaks[4]

\definecolor{cover}{rgb}{0.77,0.87,0.88}
\definecolor{blueone}{rgb}{0.1,0.1,.7}
\definecolor{citec}{rgb}{0.14,0.47,0.09}
\definecolor{two}{rgb}{0.0,0.5,0.}
\definecolor{three}{rgb}{.5,.1,0.15}

\journalname{Eur. Phys. J. C}
\begin{document}
\title{Hidden-bottom molecular states from $\Sigma^{(*)}_bB^{(*)}-\Lambda_bB^{(*)}$  interaction}
\author{Jun-Tao Zhu, Shu-Yi Kong, Yi Liu, Jun He\thanksref{e1}
}                     
\thankstext{e1}{Corresponding author: junhe@njnu.edu.cn}
\institute{Department of  Physics and Institute of Theoretical Physics, Nanjing Normal University,
Nanjing 210097, China
}
\date{Received: date / Revised version: date}
%
\maketitle
\abstract{
In this work,  we  study  possible hidden-bottom molecular pentaquarks $P_b$ from coupled-channel  $\Sigma^{(*)}_bB^{(*)}-\Lambda_bB^{(*)}$  interaction in the quasipotential Bethe-Salpeter equation approach. 
In isodoublet sector with $I=1/2$, with the same reasonable parameters the interaction  produces seven molecular states,  a state near $ \Sigma_bB$ threshold with  spin parity $J^P=1/2^-$, a state near $\Sigma^*_bB$ threshold with  $3/2^-$, two states near $\Sigma_bB^*$  threshold with  $1/2^-$ and $3/2^-$, and three states near $\Sigma_b^*B^*$ threshold with $1/2^-$, $3/2^-$, and $5/2^-$.    The results suggest that three states near $\Sigma_b^* B^*$  threshold and two states near $\Sigma_b B^*$ threshold  are very close, respectively, which may be difficult to distinguish in  experiment without partial wave analysis. Compared with the hidden-charm pentaquark, the $P_b$ states  are relatively narrow with widths at an order of magnitude of 1 MeV or smaller.  The importance of each channel considered is also discussed, and it is found that the  $\Lambda_b B^*$ channel  provides  important contribution for the widths of those states. In  isoquartet sector with $I=3/2$,  cutoff should be considerably enlarged to achieve  bound states from the interaction, which  makes the existence of such states unreliable.  The results in the current work are helpful  for searching for hidden-bottom molecular pentaquarks in  future experiments, such as the COMPASS, J-PARC, and the Electron Ion Collider  in China (EicC).}

\section{INTRODUCTION}

It is one of the most important topic in  hadron physics community to search for the hadronic exotic states beyond the conventional quark model.  Among the theoretical pictures in the market, molecular state is a competitive one to explain existing candidates of exotic states, such as the $XYZ$ particles and $P_c$ states~\cite{Chen:2016qju}. A molecular state is analogous to a nucleus, especially the deuteron, that is, a loosely bound state of two or more hadrons. It immediately leads to a conclusion that a molecular state is close to the threshold of constituent hadrons. In practice, the study of the molecular state also focuses on  resonance structures near  thresholds. Vise versa, if we can find more structures near thresholds, especially those with corresponding relationship, it will strongly support  existence of  molecular states.  In the current work, we will provide  predictions of  hidden-charm pentaquarks $P_b$, which are  partners of the hidden-charm $P_c$ states.

The observation of  hidden-charm pentaquarks at LHCb is a great breakthrough of  the study of exotic states~\cite{Aaij:2019vzc,Aaij:2015tga}. It is also an important support on the molecular state picture.  Three narrow resonance structures were reported at LHCb in an update measurement as $P_c(4457)$ and $P_c(4440)$ states near $\Sigma_c \bar{D}^*$ threshold and a $P_c(4312)$ state near $\Sigma_c\bar{D}$ threshold~\cite{Aaij:2019vzc}. Combined with $P_c(4380)$ near  $\Sigma^*_c\bar{D}$ threshold suggested in the first observation~\cite{Aaij:2015tga}, it exhibits a good pattern of the S-wave molecular states from interactions corresponding to the thresholds. Such observation confirms  the prediction of existence of the hidden-charm pentaquark in some models~\cite{Wu:2010jy,Wang:2011rga,Yang:2011wz,Wu:2012md,Yuan:2012wz}

A lot of  theoretical interpretations of these structures emerged after the experimental observation.  Due to the strong correlation between these structures and the thresholds,  the molecular state  is the most popular  picture to explain the $P_c$ states\cite{Chen:2015loa,Chen:2015moa,Karliner:2015ina,Roca:2015dva,He:2015cea,He:2016pfa,Chen:2019asm,Fernandez-Ramirez:2019koa,Huang:2018wed,Wang:2019got}, though other interpretations can not be excluded~\cite{Lebed:2015tna,Meissner:2015mza,Burns:2015dwa,Chen:2019bip}. In Ref.~\cite{Liu:2019tjn}, authors even proposed  existence of seven hidden-charm molecular states as a complete heavy-quark spin symmetry multiplet. 
In our previous works~\cite{He:2019ify,He:2019rva}, we systematically investigate  coupled-channel $\Sigma^{(*)}_c\bar{D}^{(*)}-\Lambda_c\bar{D}^{(*)}$  interaction.  Three isodoublet states with $I=1/2$ are produced near  $\Sigma_c\bar{D}$ threshold with spin parity $J^P=1/2^-$ and $\Sigma_c\bar{D}^*$ threshold with  $1/2^-$ and $ 3/2^-$. Their masses and widths fall well in the ranges of  experimental values of the $P_c(4312)$, $P_c(4440)$ and $P_c(4457)$  observed at LHCb.  A state almost on the $\Sigma_c^*\bar{D}^*$ threshold with $3/2^-$ is also produced and can be related to the $P_c(4380)$.  Two  another states near $\Sigma_c^*\bar{D}^*$ threshold with $1/2^-$ and $3/2^-$  were also produced with the same parameters, but the result suggests that their effects on the experimental observable may be small.  Besides, the decay pattern was also discussed,  and  the $\Lambda\bar{D}^*$ channel is found  dominant in the decays of these states.

Now that  experimentally observed $P_c$ states were well interpreted in the molecular state picture, we can predict hidden-bottom states  above 11 GeV.  Though there are a large amount of works about $P_c$ states reported in the literature,  the studies about the hidden-bottom pentaquark are still inadequate.  There are a few incidental studies about $P_b$ states in the works to interprete the $P_c$ states in the molecular picture,  such as within constituent quark model~\cite{Huang:2018wed,Yang:2018oqd}, the chiral effective field theory~\cite{Wang:2019ato,Chen:2015loa}, and the Bethe-Salpeter equation~\cite{Ke:2019bkf}. The width and decay pattern of the $P_b$ states were also discussed in  Refs.~\cite{Huang:2018wed,Lin:2018kcc,Lin:2019qiv,Gutsche:2019mkg,Ke:2019bkf}.  Due to  very large mass of the hidden-bottom pentaquark, it is relatively difficult to search for the   $P_b$ state in experiment compared with the $P_c$ state. In Ref.~\cite{Wang:2019zaw},  pion and photon induced productions  of hidden-bottom pentaquarks were studied, the calculation suggests that it is possible to search for these states at COMPASS, J-PARC and EicC. Hence, it is interesting to perform a systematical study about the hidden-charm pentaquarks based on  experimental information and  theoretical analysis about the $P_c$ states.

In this work, we will  investigate  coupled-channel $\Sigma^{(*)}_bB^{(*)}-\Lambda_bB^{(*)}$ interaction in the quasipotential Bethe-Salpeter equation (qBSE) approach to find possible hidden-bottom molecular states.  The interaction was described in the one-boson-exchange model with the help of the effective Lagrangians within the heavy quark symmetry and chiral limit  as in Refs.~\cite{He:2019ify,He:2019rva}, where the $P_c$ states were interpreted. The masses and widths of molecular states are predicted  by finding  poles in  complex energy plane. The decay channels of predicted states will be discussed also.

 This article is organized as follows. After introduction, the details of theoretical frame of coupled-channel $\Sigma^{(*)}_bB^{(*)}-\Lambda_bB^{(*)}$ interactions is presented in section~\ref{Sec: Formalism}.  In Section~\ref{3}, the single-channel  results of the states with isospin $I=1/2$ and $I=3/2$ are given first.   Then,  coupled-channel results are presented, and the importance of the channels considered  are discussed. Finally, summary  and  discussion will be given in section~\ref{5}.

\section{Theoretical frame}\label{Sec: Formalism}

In the qBSE approach, we will use the one-boson-exchange interaction of two bottom hadrons as  dynamical kernel. In the current work, we will adopt the Lagrangians with heavy quark limit and chiral symmetry, and the channels with hidden-charm mesons are ignored as in Ref.~\cite{He:2019ify,He:2019rva} to keep the consistence.  The peseudoscalar $\mathbb{P}$, vector $\mathbb{V}$ and scalar $\sigma$  exchanges  will be considered, and the effective Lagrangians depicting the couplings of light mesons and  bottom mesons or bottom baryons  are required and will be presented in the below.

First, we consider the couplings of light mesons to heavy-light bottom mesons $\mathcal{P}=(B^0, B^+, B^+_s)$. The Lagragians were  constructed  in the literature as \cite{Cheng:1992xi,Yan:1992gz,Wise:1992hn,Casalbuoni:1996pg},
\begin{align} 
  \mathcal{L}_{\mathcal{P}^*\mathcal{P}\mathbb{P}} &=
 i\frac{2g\sqrt{m_{\mathcal{P}} m_{\mathcal{P}^*}}}{f_\pi}
  (-\mathcal{P}^{*\dag}_{a\lambda}\mathcal{P}_b
  +\mathcal{P}^\dag_{a}\mathcal{P}^*_{b\lambda})
  \partial^\lambda\mathbb{P}_{ab},\nonumber\\
    \mathcal{L}_{\mathcal{P}^*\mathcal{P}^*\mathbb{P}} &=
-\frac{g}{f_\pi} \epsilon_{\alpha\mu\nu\lambda}\mathcal{P}^{*\mu\dag}_a
\overleftrightarrow{\partial}^\alpha \mathcal{P}^{*\lambda}_{b}\partial^\nu\mathbb{P}_{ba},\nonumber\\
    \mathcal{L}_{\mathcal{P}^*\mathcal{P}\mathbb{V}} &=
\sqrt{2}\lambda g_V\varepsilon_{\lambda\alpha\beta\mu}
  (-\mathcal{P}^{*\mu\dag}_a\overleftrightarrow{\partial}^\lambda
  \mathcal{P}_b  +\mathcal{P}^\dag_a\overleftrightarrow{\partial}^\lambda
 \mathcal{P}_b^{*\mu})(\partial^\alpha{}\mathbb{V}^\beta)_{ab},\nonumber\\
	\mathcal{L}_{\mathcal{P}\mathcal{P}\mathbb{V}} &= -i\frac{\beta	g_V}{\sqrt{2}}\mathcal{P}_a^\dag
	\overleftrightarrow{\partial}_\mu \mathcal{P}_b\mathbb{V}^\mu_{ab}, \nonumber\\
  \mathcal{L}_{\mathcal{P}^*\mathcal{P}^*\mathbb{V}} &= - i\frac{\beta
  g_V}{\sqrt{2}}\mathcal{P}_a^{*\dag}\overleftrightarrow{\partial}_\mu
  \mathcal{P}^*_b\mathbb{V}^\mu_{ab}-i2\sqrt{2}\lambda  g_Vm_{\mathcal{P}^*}\mathcal{P}^{*\mu\dag}_a\mathcal{P}^{*\nu}_b\mathbb{V}_{\mu\nu ab}
,\nonumber\\
  \mathcal{L}_{\mathcal{P}\mathcal{P}\sigma} &=
  -2g_s m_{\mathcal{P}}\mathcal{P}_a^\dag \mathcal{P}_a\sigma, \nonumber\\
  \mathcal{L}_{\mathcal{P}^*\mathcal{P}^*\sigma} &=
  2g_s m_{\mathcal{P}^*}\mathcal{P}_a^{*\dag}
  \mathcal{P}^*_a\sigma,\label{LD}
\end{align}
where  $f_\pi=132$ MeV, $\mathbb{V}_{\mu\nu}=\partial_\mu\mathbb{V}_\nu-\partial_\nu\mathbb{V}_\mu$. The
$\mathcal{P}
$ and $\mathcal{P}^*
$ satisfy the normalization relations $\langle
0|{\mathcal{P}}|\bar{Q}{q}(0^-)\rangle
=\sqrt{M_\mathcal{P}}$ and $\langle
0|{\mathcal{P}}^*_\mu|\bar{Q}{q}(1^-)\rangle=
\epsilon_\mu\sqrt{M_{\mathcal{P}^*}}$.
The $\mathbb
P$ and $\mathbb V$ are the pseudoscalar and vector matrices as
\begin{align}
    {\mathbb P}&=\left(\begin{array}{ccc}
        \frac{\sqrt{3}\pi^0+\eta}{\sqrt{6}}&\pi^+&K^+\\
        \pi^-&\frac{-\sqrt{3}\pi^0+\eta}{\sqrt{6}}&K^0\\
        K^-&\bar{K}^0&-\frac{2\eta}{\sqrt{6}}
\end{array}\right),
\mathbb{V}=\left(\begin{array}{ccc}
\frac{\rho^0+\omega}{\sqrt{2}}&\rho^{+}&K^{*+}\\
\rho^{-}&\frac{-\rho^{0}+\omega}{\sqrt{2}}&K^{*0}\\
K^{*-}&\bar{K}^{*0}&\phi
\end{array}\right).\label{MPV}
\end{align}

The explicit forms of the Lagrangians for the couplings of light mesons to bottom baryons can be written as~\cite{Liu:2011xc},
\begin{align}
{\cal L}_{BB\mathbb{P}}&=i\frac{3g_1}{2f_\pi\sqrt{m_{\bar{B}}m_{B}}}~\epsilon^{\mu\nu\lambda\kappa}\partial^\nu \mathbb{P}~
\sum_{i=0,1}\bar{B}_{i\mu} \overleftrightarrow{\partial}_\kappa B_{j\lambda},\nonumber\\
{\cal L}_{BB\mathbb{V}}&=-\frac{\beta_S g_V}{\sqrt{2m_{\bar{B}}m_{B}}}\mathbb{V}^\nu
 \sum_{i=0,1}\bar{B}_i^\mu \overleftrightarrow{\partial}_\nu B_{j\mu}-\frac{\lambda_S
g_V}{\sqrt{2}}\mathbb{V}_{\mu\nu}\sum_{i=0,1}\bar{B}_i^\mu B_j^\nu,\nonumber\\
    {\cal L}_{B_{\bar{3}}B_{\bar{3}}\mathbb{V}}&=-\frac{g_V\beta_B}{\sqrt{2m_{\bar{B}_{\bar{3}}}m_{B_{\bar{3}}}} }\mathbb{V}^\mu\bar{B}_{\bar{3}}\overleftrightarrow{\partial}_\mu B_{\bar{3}},\nonumber\\
{\cal L}_{BB_{\bar{3}}\mathbb{P}}
    &=-i\frac{g_4}{f_\pi} \sum_i\bar{B}_i^\mu \partial_\mu \mathbb{P} B_{\bar{3}}+{\rm H.c.},\nonumber\\
{\cal L}_{BB_{\bar{3}}\mathbb{V}}    &=\sqrt{2\over m_{\bar{B}}m_{B_{\bar{3}}}}{g_\mathbb{V}\lambda_I} \epsilon^{\mu\nu\lambda\kappa} \partial_\lambda \mathbb{V}_\kappa\sum_i\bar{B}_{i\nu} \overleftrightarrow{\partial}_\mu
   B_{\bar{3}}+{\rm H.c.}.\nonumber\\
 {\cal L}_{BB\sigma}&=\ell_S\sigma\sum_{i=0,1}\bar{B}_i^\mu B_{j\mu},\nonumber\\
{\cal L}_{B_{\bar{3}}B_{\bar{3}}\sigma}&=i\ell_B \sigma \bar{B}_{\bar{3}}B_{\bar{3}},
   \label{LB}
\end{align}
where $B_{i\mu}$ is defined as 
\begin{align}
    \left({ B}^{ab}_{0\mu}, B^{ab}_{1\mu}\right)&\equiv \left(-\sqrt{\frac{1}{3}}(\gamma_{\mu}+v_{\mu})
    \gamma^{5}B^{ab}, B^{*ab}_{\mu}\right),
\end{align}
and the bottomed baryon matrices are defined as
\begin{align}
B_{\bar{3}}&=\left(\begin{array}{ccc}
0&\Lambda^+_b&\Xi_b^+\\
-\Lambda_b^+&0&\Xi_b^0\\
-\Xi^+_b&-\Xi_b^0&0
\end{array}\right),\
B=\left(\begin{array}{ccc}
\Sigma_b^{++}&\frac{1}{\sqrt{2}}\Sigma^+_b&\frac{1}{\sqrt{2}}\Xi'^+_b\\
\frac{1}{\sqrt{2}}\Sigma^+_b&\Sigma_b^0&\frac{1}{\sqrt{2}}\Xi'^0_b\\
\frac{1}{\sqrt{2}}\Xi'^+_b&\frac{1}{\sqrt{2}}\Xi'^0_b&\Omega^0_b
\end{array}\right).\label{MBB}
\end{align}

In the calculation,  the masses of  particles  are chosen as suggested central values in the Review of  Particle Physics  (PDG)~\cite{Tanabashi:2018oca}. The mass of broad $\sigma$ meson is chosen as 500 MeV.  The  coupling constants involved was cited from the literature~\cite{Chen:2019asm,Liu:2011xc,Isola:2003fh,Falk:1992cx}, and listed in Table~\ref{coupling}, 
\vspace{-1.5em}
\renewcommand\tabcolsep{0.13cm}
\renewcommand{\arraystretch}{1.2}
\begin{table}[h!]
\caption{The coupling constants adopted in our
calculation. The $\lambda$ and $\lambda_{S,I}$ are in the unit of GeV$^{-1}$. Others are in the unit of $1$.
\label{coupling}}
\begin{tabular}{cccccccccccccccccc}\bottomrule[2pt]
$\beta$&$g$&$g_V$&$\lambda$ &$g_{s}$\\
0.9&0.59&5.9&0.56 &0.76\\\hline
$\beta_S$&$\ell_S$&$g_1$&$\lambda_S$ &$\beta_B$&$\ell_B$ &$g_4$&$\lambda_I$\\
-1.74&6.2&-0.94&-3.31&$-\beta_S/2$&$-\ell_S/2$&$3g_1/{(2\sqrt{2})}$&$-\lambda_S/\sqrt{8}$ \\
\bottomrule[2pt]
\end{tabular}
\end{table}
\vspace{-1.5em}

With the vertices obtained from the  above Lagrangians, the potential of couple-channel   interaction can be constructed.  
Because six channels are invovled in the current work, it is
tedious and fallible  to give explicit 36 potential elements and input them into  code.  Instead, in this work, we input   vertices $\Gamma$ and  propagators $P$  into  code directly, and the potential can be obtained as
\begin{equation}%
{\cal V}_{\mathbb{P},\sigma}=f_I\Gamma_1\Gamma_2 P_{\mathbb{P},\sigma}f(q^2),\ \ 
{\cal V}_{\mathbb{V}}=f_I\Gamma_{1\mu}\Gamma_{2\nu}  P^{\mu\nu}_{\mathbb{V}}f(q^2),\label{V}
\end{equation}
The propagators are defined as usual as
\begin{equation}%
P_{\mathbb{P},\sigma}= \frac{i}{q^2-m_{\mathbb{P},\sigma}^2},\ \
P^{\mu\nu}_\mathbb{V}=i\frac{-g^{\mu\nu}+q^\mu q^\nu/m^2_{\mathbb{V}}}{q^2-m_\mathbb{V}^2},
\end{equation}
where the form factor $f(q^2)$ is adopted to compensate the off-shell effect of exchanged meson as $f(q^2)=e^{-(m_e^2-q^2)^2/\Lambda_e^2}$
with $m_e$ being the $m_{\mathbb{P},\mathbb{V},\sigma}$ and $q$ being the momentum of the exchanged  meson. The cutoff is rewritten as a form of $\Lambda_e=m+\alpha_e~0.22$ GeV.
The $f_I$ is the flavor factor for certain meson exchange of certain interaction, and the explicit values are listed in Table~\ref{flavor factor}.
\vspace{-1.5em}
\renewcommand\tabcolsep{0.27cm}
\renewcommand{\arraystretch}{1.5}
\begin{table}[h!]
\caption{The flavor factors $f_I$ for certain meson exchanges of certain interaction. The values in bracket are for the case of $I=3/2$ if the values are different from these of $I=1/2$.  \label{flavor factor}}
{	\begin{tabular}{c|ccccc}\bottomrule[2pt]
& $\pi$&$\eta$&$\rho$ & $\omega$ & $\sigma$ \\\hline
$B^{(*)}\Sigma^{(*)}_b\to B^{(*)}\Sigma^{(*)}_b$&$-1[\frac{1}{2}]$ &$\frac{1}{6}[\frac{1}{6}]$ &$-1[\frac{1}{2}]$&$\frac{1}{2}[\frac{1}{2}]$ & 1\\
$B^{(*)}\Lambda_b\to B^{(*)}\Lambda_b$&$0$ &$0$ &$0$&$1$ & 2\\
$B^{(*)}\Lambda_b\to B^{(*)}\Sigma_b^{(*)}$&$\sqrt{6}\over{2}$ &$0$ &$\sqrt{6}\over{2}$&$0$ & 0\\
\toprule[2pt]
\end{tabular}}
\end{table}
\vspace{-1.5em}

With the potential kernel obtained, we use the qBSE to solve the scattering amplitude~\cite{He:2014nya,He:2015mja,He:2012zd,He:2015yva,He:2017aps}. After partial-wave decomposition and spectator quasipotential approximation,  the 4-dimensional Bethe-Saltpeter equation  in the Minkowski space can be reduced to a 1-dimensional  equation with fixed spin-parity $J^P$ as~\cite{He:2015mja},
\begin{align}
i{\cal M}^{J^P}_{\lambda'\lambda}({\rm p}',{\rm p})
&=i{\cal V}^{J^P}_{\lambda',\lambda}({\rm p}',{\rm
p})+\sum_{\lambda''}\int\frac{{\rm
p}''^2d{\rm p}''}{(2\pi)^3}\nonumber\\
&\cdot
i{\cal V}^{J^P}_{\lambda'\lambda''}({\rm p}',{\rm p}'')
G_0({\rm p}'')i{\cal M}^{J^P}_{\lambda''\lambda}({\rm p}'',{\rm
p}),\quad\quad \label{Eq: BS_PWA}
\end{align}
where the sum extends only over nonnegative helicity $\lambda''$.
Here, the
reduced propagator with the spectator approximation can be written as  $G_0({\rm p}'')=\delta^+(p''^{~2}_h-m_h^{2})/(p''^{~2}_l-m_l^{2})$ with $p''_{h,l}$ and $m_{h,l}$ being the momenta and masses of heavy or light constituent particles.
The partial wave potential is defined with the potential of  interaction obtained in the above in Eq.~(\ref{V}) as
\begin{align}
{\cal V}_{\lambda'\lambda}^{J^P}({\rm p}',{\rm p})
&=2\pi\int d\cos\theta
~[d^{J}_{\lambda\lambda'}(\theta)
{\cal V}_{\lambda'\lambda}({\bm p}',{\bm p})\nonumber\\
&+\eta d^{J}_{-\lambda\lambda'}(\theta)
{\cal V}_{\lambda'-\lambda}({\bm p}',{\bm p})],
\end{align}
where $\eta=PP_1P_2(-1)^{J-J_1-J_2}$ with $P$ and $J$ being parity and spin for system, $B^{(*)}$ meson or $\Sigma_b^{(*)}$ baryon. The initial and final relative momenta are chosen as ${\bm p}=(0,0,{\rm p})$  and ${\bm p}'=({\rm p}'\sin\theta,0,{\rm p}'\cos\theta)$. The $d^J_{\lambda\lambda'}(\theta)$ is the Wigner d-matrix.
we also adopt an  exponential
regularization  by introducing a form factor into the propagator as~\cite{He:2015mja}
\begin{equation}
G_0({\rm p}'')\to G_0({\rm p}'')\left[e^{-(p''^2_l-m_l^2)^2/\Lambda_r^4}\right]^2.\label{regularization}
\end{equation}
 In the current work,  the relation of the cutoff $\Lambda_r=m+\alpha_r~0.22$ GeV with $m$ being the mass of the exchanged meson is also introduced into the regularization form factor to suppress  large  momentum, $i.\ e.$, the short-range contribution of the $\pi$ exchange as warned in Ref.~\cite{Liu:2019zvb}.

\section{Numerical results}\label{3}

The 1-dimensional integral equation can be transformed into a matrix equation as $M=V+VG_0M$ by Gauss discretization. The molecular states can be found by searching for the pole of scattering  amplitude $ M$ in complex energy plane at $|1-V(z)G(z)|=0$ with $z=W+i\Gamma/2$ equaling to  system energy $W$ at 
real axis~\cite{He:2015mja}.  In addition, we take two free parameters $\alpha_e$ and $\alpha_r$ as $\alpha$ for simplification.

\subsection{Single-channel results}


Each experimental observed $P_c$ state is close to a threshold, respectively~\cite{Aaij:2019vzc,Aaij:2015tga}.  It suggests that each of these states should be mainly from a single-channel interaction in the molecular state picture, which is confirmed by previous study in Ref.~\cite{He:2019rva}. In this work, we  present the single-channel results first. 

\begin{figure}[h!]
  \flushleft
  \includegraphics[bb=40 40 1000 840, scale=0.35] {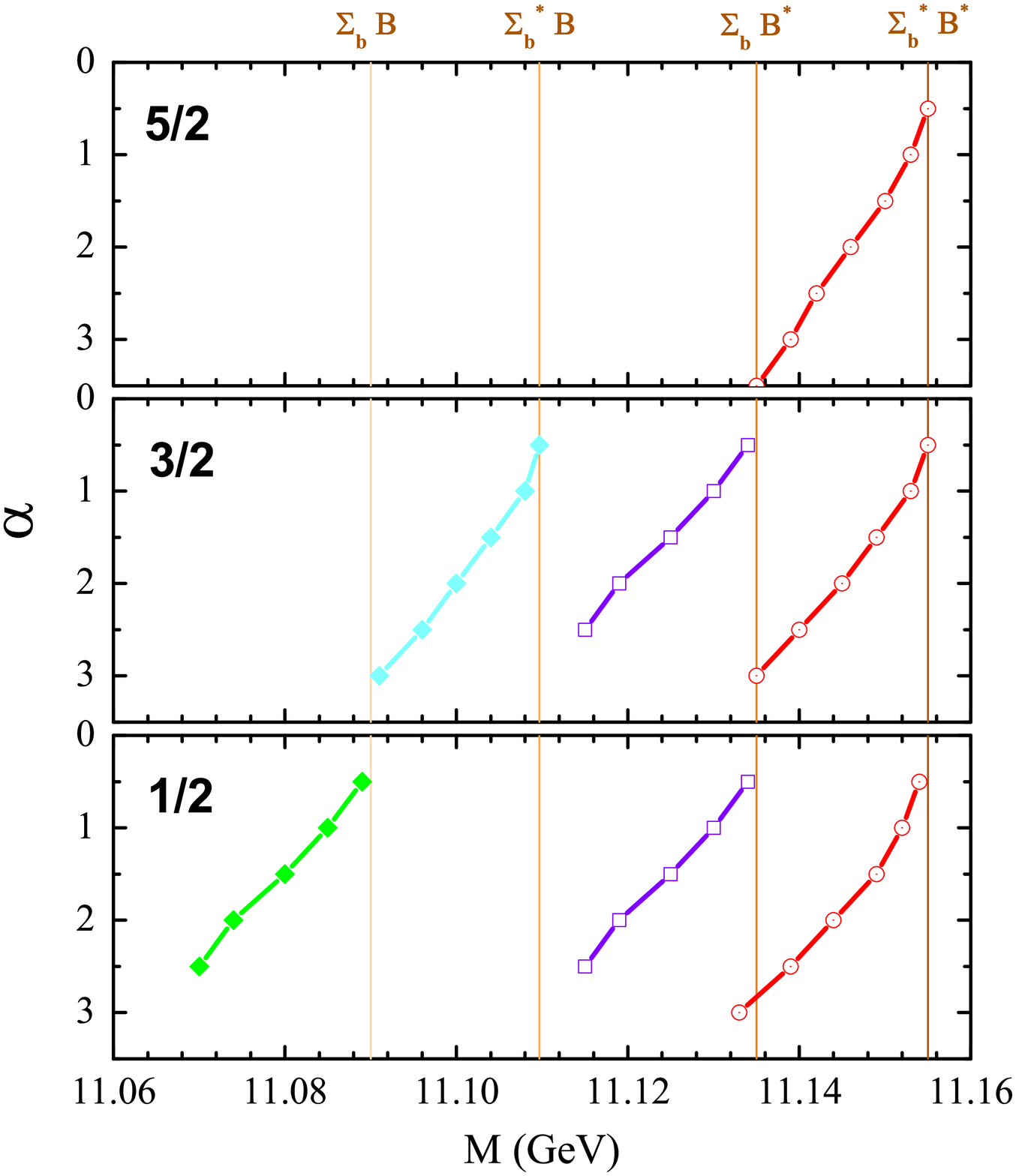}
  \caption{The  $\alpha$ dependence of the mass $M$ of isodoublet binding states from  single-channel interaction. The four solid lines from right to left represent the thresholds of four channels $\Sigma_b^* B^*$, $\Sigma_b B^*$, $\Sigma_b^* B$ and $\Sigma_bB$ at 11155MeV, 11135MeV, 11110MeV and 11090MeV, respectively. The curves  are for the bound states from the interactions with corresponding thresholds.}
  \label{massI1}
\end{figure}

In the current work, we consider all states with spin parities which can be produced from S-wave interaction, $\Sigma_b^* B^*$ with $1/2^{-}, 3/2^{-}, 5/2^{-}$, $\Sigma_b B^*$ with $1/2^{-}, 3/2^{-}$, $\Sigma_b^* B$ with $3/2^{-}$ and $\Sigma_b B$ with $1/2^-$. The results for isodoublet with $I=1/2$ are illustrated in  Fig.~\ref{massI1}. The $\Lambda_bB^{(*)}$ with $1/2^-, 3/2^-$ and  $\Lambda_bB$ with $1/2^-$ are also calculated, however,  large $\alpha$ beyond reasonable limit is required to produce  bound states.
\vspace{-1.5em}
\vspace{-1.5em}

The results suggest that  bound states can be produced in all seven cases in a range of $\alpha$ from 0 to 3.5.  All states appear at $\alpha$ about 0.5,  and binding becomes deeper with the increase of  $\alpha$, and gradually reach to a binding energy  about 20 MeV or the next threshold at $\alpha=2.5-3.5$ . The binding energies will continue to increase, but we no longer present such results.   The trends of curves for three states produced from the $\Sigma_b^*B^*$ interaction with different spin parities are almost the same. Such phenomenon can also be found for  two curves for two states from the $\Sigma_bB^*$ interaction. 
Compared with the results for $P_c$ states~\cite{He:2019ify,He:2019rva},  one can find that the values of  parameter $\alpha$ to produce the hidden-bottom molecular states is relatively smaller.

\begin{figure}[htpb!]
\flushleft
\includegraphics[bb=40 40 1000 840, scale=0.35]{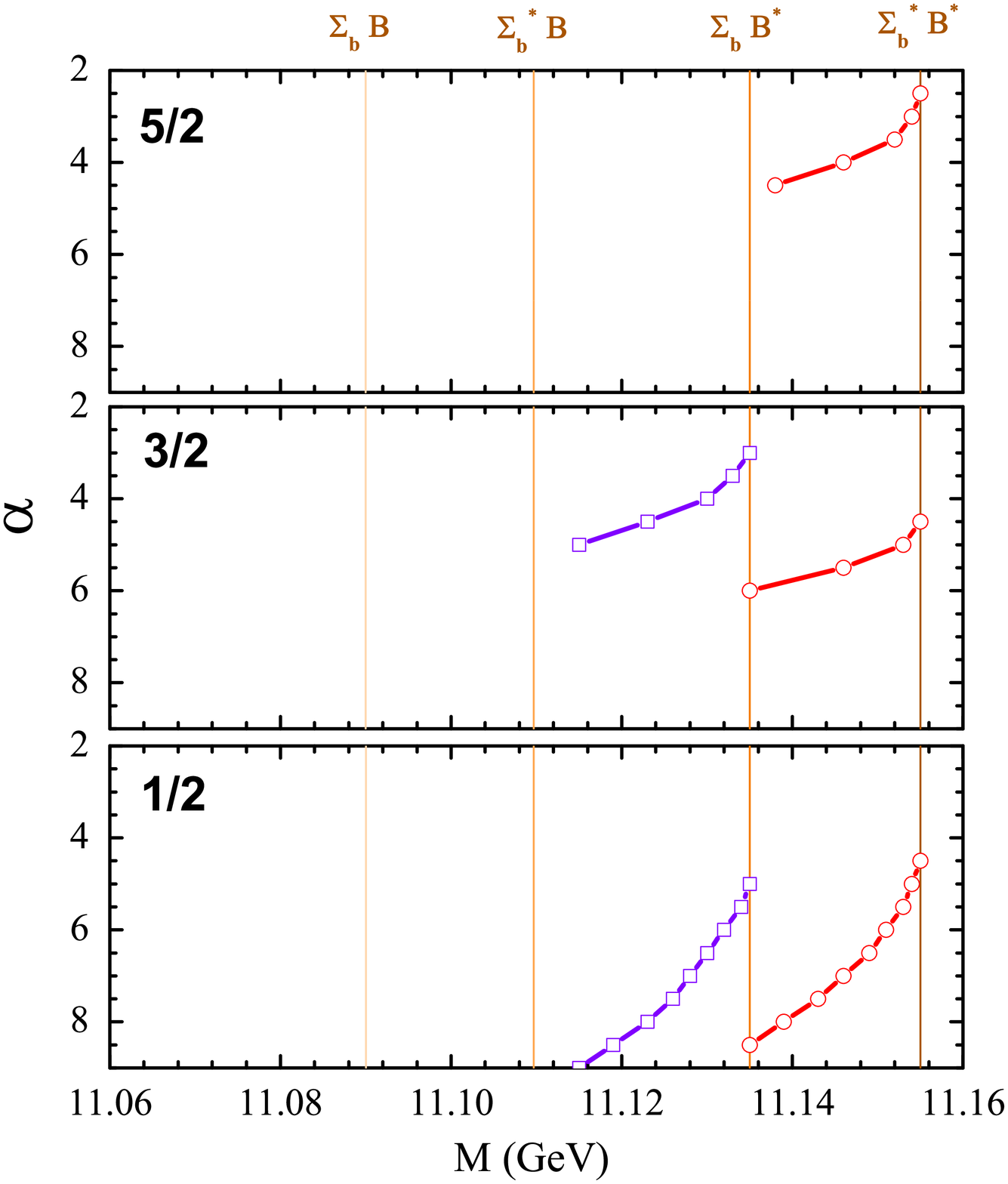}
\caption{The mass $M$ with the variation of the $\alpha$ for isoquartet bound states. Other conventions are the same as in Fig. \ref{massI1}.}
\label{massI3}
\end{figure}

In Fig~\ref{massI3}, we present the results for isoquartet states with $I=3/2$. In the calculation, we also consider seven cases as for isodoublet.  No bound state is produced form $\Sigma_bB$ interaction with $(1/2^-)$ and $\Sigma_b^*B$ interaction from $(3/2^-)$  even if $\alpha$  is taken to 9.
Except these two states, left five states can be produced from  single-channel interaction as shown in ~\ref{massI3}, but  with considerablly large $\alpha$. The production of bound states, $\Sigma_b^*B^*(1/2^-)$, $\Sigma_bB^*(1/2^-)$ and $\Sigma_b^*B^*(3/2^-)$,  needs a value of   $\alpha$ at least 4, which is  larger than the maximum value of $\alpha$ required for  binding of isodoublet molecular states with $I=1/2$. It indicates that these three states are hardly to be found if the isoboublet states have small binding energies.
The rest two states  $\Sigma_bB^*(3/2^-) $ and $\Sigma_b^*B^*(5/2^-)$ appears at $\alpha=3$ and $2.5$,  respectively. it implies that these two molecular states may exist if the isoboublet states are deeply bound.  Generally speaking, if we assume that the $P_b$ states are also loosely bound states as $P_c$ states, the possibility of existence of  isoquaret states is vey small.   In Ref~\cite{Yang:2018oqd}, within the frame of constituent quark model, the molecular states with  $I=3/2$ were also not found.
\vspace{-1.5em}
\vspace{-3.em}

\subsection{Coupled-channel results}\label{4}

In the previous calculation, the bound states from single channel calculation exhibit as poles at  real axis of  complex energy plane, that is, the widths of these states are zero. In the case of the $P_c$ states, we found that the experimental width can be well reproduced with inclusion of  coupled-channel effect~\cite{He:2019rva}. 
In the above single-channel calculation, seven bound states are produced. Those states can be coupled to each other by exchanges of light mesons.  Besides, the $\Lambda\bar{D}^{*}$ channel is also found important for  the width of  $P_c$ states~\cite{He:2019rva,Lin:2018kcc,Lin:2019qiv}. In the following,  coupled-channel results for the $\Sigma^{(*)}_bB^{(*)}-\Lambda_bB^{(*)}$ interaction will be given.

\begin{figure}[h!]
\flushleft
\includegraphics[bb=80 45 500 300, scale=0.81]{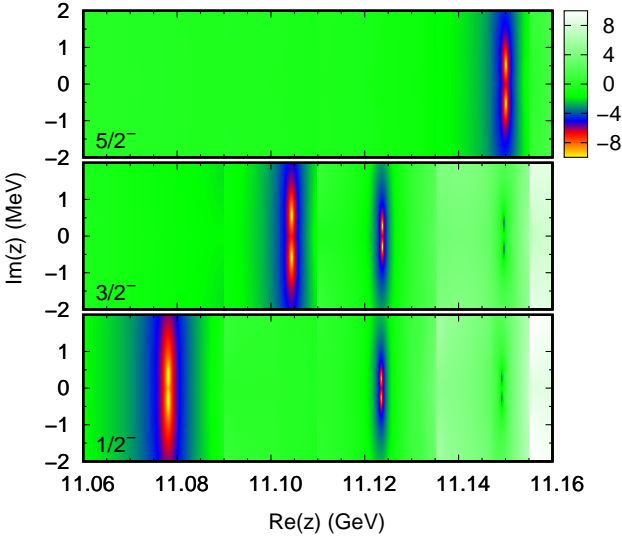}
\caption{The $\log|1-V(z)G_0(z)|$ with the variation of  $z$ for coupled-channel $\Sigma^{(*)}_bB^{(*)}-\Lambda_bB^{(*)}$ interaction with $J^P=1/2^-$, $3/2^-$and $5/2^-$  at $\alpha=1.5$. The  color  means the value of $\log|1-V(z)G_0(z)|$ as shown in the color box.}
\label{cc1}
\end{figure}

Here, we first give an example to show a general picture of  coupled-channel results. Since there is no experimental data about the $P_b$ states, we should choose a parameter to present the results.  The only free parameter in our model is $\alpha$, and in the single-channel calculation the  $\alpha$ dependences of the masses of seven bound states exhibit a similar trend.  Hence, we choose $\alpha$ as  1.5 to illustrate the poles from  coupled-channel $\Sigma^{(*)}_bB^{(*)}-\Lambda_bB^{(*)}$ interaction  in Fig.~\ref{cc1}. The values of $\log|1-V(z)G_0(z)|$  with variation of complex energy $z$ is adopted to show the positions of poles of coupled-channel scattering amplitude because $M=(1-VG_0)^{-1}V$.  And we present the results for spin parities $5/2^-$, $3/2^-$, and $1/2^-$ in a range from 11.06  to 11.16 GeV for real part of complex energy $Re(z)$ and -2  to 2 MeV for imaginary part $Im(z)$. 
\vspace{-1.5em}
\vspace{-1.5em}

One can find that there are still seven poles produced  as in the single-channel calculation. It suggests that only states with an S-wave interaction can be produced for  three spin parities considered.  In the case with $J^P=1/2^-$, there exist three poles near the $\Sigma_bB$ (S wave), $\Sigma_bB^*$ (S and D waves) and $\Sigma^*_bB^*$ (S, D and G waves) thresholds, respectively.  No pole appears near $\Sigma^*B$ (P wave) threshold for $1/2^-$. In the case with $J^P=3/2^-$, we also have three poles near the $\Sigma^*_bB$  (S and D waves), $\Sigma_bB^*$  (S and D waves) and $\Sigma^*_bB^*$  (S, D and G waves) thresholds, respectively. And no pole appears near $\Sigma_b B$ (P wave) threshold. For spin parity $5/2^-$ there is only one pole near the  $\Sigma^*_bB^*$ thresholds, and only this channel can produce a pole with $5/2^-$ in S wave.

The $P_c(4457)$ and $P_c(4440)$ is close to each other near $\Sigma_c\bar{D}^*$ threshold, which was even taken as one resonance structure in the first observation of the $P_c$ states~\cite{Aaij:2019vzc}.  In the $P_b$ case, our results exhibit a more serious overlapping between two states near $\Sigma_bB^*$ as in the single channel calculation. A shown in Fig.~\ref{cc1}, these two poles with $1/2^-$ and $3/2^-$ have almost the same mass and width. 
Furthermore, the masses of three poles near the $\Sigma^*_bB^*$ threshold are also close very much, which are $11149.2$~MeV, $11149.7$~MeV and $11150.1$~MeV corresponding to the two molecular states with $1/2^-$, $3/2^-$ and $5/2^-$, respectively. Different from the $P_c$ states~\cite{He:2019rva}, here, the state with $5/2^-$ stands out  background obviously while  other two poles with $J^P=1/2^-$ and $3/2^-$ are very dimly, and may be difficult to be found at experiment. 

As shown in the figure, the poles acquire  imaginary parts after  coupled-channel effects are included in the calculation. However,  these states are generally very narrow, with imaginary parts smaller than 0.5~MeV. By using the relation $\Gamma=-2$ Im$(z)$, it means a small width about 1 MeV or smaller, which is much smaller than the $P_c$ states with similar binding energy. It is worth mentioning that at  $\alpha=1.5$, except for $\Sigma_b^*B^*(5/2^-)$, the masses of  six molecular states after inclusion of coupled-channel effect are in good agreement with the results obtained under the frame of the constituent quark model in Ref~\cite{Huang:2018wed}.

In the above, we only present the poles of molecular states obtained by the  coupled-channels calculation at a value of  $\alpha=1.5$. Because there is no experimental data, in the following, we will present the results with different vales of $\alpha$ from 0.5 to 2.0 to show the dependence of  results on the parameter in second and third columns of Table~\ref{pole}.  The two-channel calculation results are also listed in the fourth to  eighth columns to show the role of each channel on  widths of molecular states.  Here, to emphasize the nearest threshold, we replace the real part of  pole by $z\to M_{th}-z$ with $M_{th}$ being the mass of nearest higher threshold.

\renewcommand\tabcolsep{0.395cm}
\renewcommand{\arraystretch}{1.5}
\begin{table*}[hpbt!]   
\caption{The masses and widths of  molecular states at different values of $\alpha$. The ``$CC$" means  full coupled-channel calculation.  The values of the complex position means mass of corresponding threshold subtracted by the position of a pole, $M_{th}-z$,  in the unit of MeV. The two short line "$--$" means the coupling does not exist. The imaginary part of some poles  are shown as "$0.00$", which means too small  value under the current precision chosen.
\label{pole}}
\begin{tabular}{r|rr|rrrrrrrrr|rr}\bottomrule[2pt]
&$\alpha_r$ &\multicolumn{1}{c|}{$CC$}& \multicolumn{1}{c}{$\Sigma_b B^*$} & \multicolumn{1}{c}{$\Sigma^*_b B$} &  \multicolumn{1}{c}{ $\Sigma_b B$}&   \multicolumn{1}{c}{$\Lambda_b B^*$} &  \multicolumn{1}{c}{  $\Lambda_b B$} \\
\hline
\multirow{4}*{$ \begin{array}{c}\Sigma^*_b B^*(1/2^-) \\M_{th}=11155 {\rm MeV}\end{array}$} 
&$0.6$ &$0.6+0.02i$   &$0.6+0.01i$     &$0.6+0.01i$      &$0.6+0.01i$  &$0.6+0.00i$   &$0.6+0.00i$   \\
&$1.0$ &$2.4+0.09i$   &$2.5+0.03i$    &$2.5+0.03i$      &$2.5+0.04i$  &$2.5+0.02i$   &$2.5+0.00i$  \\
&$1.5$ &$5.8+0.28i$   &$6.1+0.07i$    &$6.0+0.10i$       &$6.0+0.16i$  &$6.0+0.16i$   &$6.1+0.00i$   \\
&$2.0$ &$9.6+0.40i$   &$10.3+0.24i$    &$10.4+0.25i$   &$9.9+0.31i$  &$10.3+0.63i$  &$10.5+0.00i$  \\
\hline
\multirow{4}*{$ \begin{array}{c}\Sigma^*_b B^*(3/2^-) \\M_{th}=11155 {\rm MeV}\end{array}$} 
&$0.6$ &$0.5+0.03i$   &$0.5+0.02i$    &$0.5+0.01i$     &$0.5+0.00i$  &$0.5+0.00i$  &$0.5+0.00i$  \\
&$1.0$ &$2.2+0.10i$   &$2.3+0.04i$   &$2.3+0.03i$      &$2.3+0.02i$  &$2.3+0.01i$   &$2.3+0.00i$  \\
&$1.5$ &$5.3+0.36i$   &$5.5+0.07i$    &$5.7+0.03i$   &$5.6+0.08i$  &$5.6+0.12i$  &$5.6+0.03i$   \\
&$2.0$ &$8.6+1.38i$   &$9.3+0.19i$     &$9.8+0.09i$     &$9.8+0.16i$  &$9.4+0.48i$   &$9.7+0.17i$   \\
\hline
\multirow{4}*{$ \begin{array}{c}\Sigma^*_b B^*(5/2^-) \\M_{th}=11155 {\rm MeV}\end{array}$} 
&$0.6$ &$0.4+0.04i$   &$0.4+0.01i$    &$0.4+0.01i$     &$0.4+0.01i$ &$0.4+0.00i$ &$0.4+0.00i$   \\
&$1.0$ &$2.1+0.15i$   &$2.4+0.01i$    &$2.1+0.05i$    &$2.0+0.01i$  &$2.1+0.01i$   &$2.1+0.00i$   \\
&$1.5$ &$4.9+0.52i$   &$5.3+0.01i$    &$5.1+0.02i$     &$5.0+0.27i$  &$4.9+0.25i$   &$5.0+0.07i$   \\
&$2.0$ &$8.6+1.38i$   &$9.3+0.19i$    &$9.8+0.09i$    &$9.8+0.16i$  &$9.4+0.48i$   &$9.7+0.17i$   \\
\hline
\multirow{4}*{$ \begin{array}{c}\Sigma_b B^*(1/2^-) \\M_{th}=11135 {\rm MeV}\end{array}$} 
&$0.5$ &$1.4+0.01i$  &\multicolumn{1}{c}{$--$}   &$1.2+0.00i$     &$1.2+0.00i$  &$1.2+0.00i$  &$1.2+0.00i$  \\
&$1.0$ &$5.7+0.05i$  &\multicolumn{1}{c}{$--$}    &$5.0+0.01i$    &$5.0+0.02i$   &$5.0+0.01i$   &$5.0+0.00i$ \\
&$1.5$ &$11.4+0.26i$ &\multicolumn{1}{c}{$--$}    &$10.0+0.05i$   &$10.1+0.05i$   &$10.0+0.07i$   &$10.1+0.03i$  \\
&$2.0$ &$17.6+0.70i$ &\multicolumn{1}{c}{$--$}  &$15.7+0.25i$    &$16.1+0.09i$   &$15.7+0.22i$   &$15.9+0.03i$  \\
\hline
\multirow{4}*{$ \begin{array}{c}\Sigma_b B^*(3/2^-) \\M_{th}=11135 {\rm MeV}\end{array}$} 
&$0.5$ &$1.4+0.02i$  &\multicolumn{1}{c}{$--$}    &$1.2+0.00i$    &$1.2+0.00i$   &$1.2+0.00i$  &$1.2+0.00i$  \\
&$1.0$ &$5.7+0.17i$  &\multicolumn{1}{c}{$--$}   &$5.1+0.01i$   &$5.1+0.14i$   &$5.1+0.02i$  &$5.1+0.00i$  \\
&$1.5$ &$11.2+0.28i$ &\multicolumn{1}{c}{$--$}    &$10.1+0.02i$  &$10.3+0.22i$   &$10.0+0.20i$   &$10.1+0.05i$  \\
&$2.0$ &$17.2+0.45i$ &\multicolumn{1}{c}{$--$}   &$15.7+0.03i$   &$16.2+0.33i$  &$15.0+0.81i$  &$15.5+0.31i$   \\
\hline
\multirow{4}*{$ \begin{array}{c}\Sigma_b^* B(3/2^-) \\M_{th}=11110 {\rm MeV}\end{array}$} 
&$1.0$ &$2.4+0.08i$  &\multicolumn{1}{c}{$--$}  &\multicolumn{1}{c}{$--$}  &$2.4+0.00i$   &$2.4+0.07i$   &$2.4+0.00i$ \\
&$1.5$ &$5.5+0.57i$  &\multicolumn{1}{c}{$--$}  &\multicolumn{1}{c}{$--$} &$5.7+0.00i$    &$5.4+0.49i$   &$5.7+0.00i$  \\
&$2.0$ &$8.4+2.05i$  &\multicolumn{1}{c}{$--$} &\multicolumn{1}{c}{$--$}  &$9.6+0.00i$   &$8.8+1.56i$   &$9.6+0.00i$ \\
\hline
\multirow{4}*{$ \begin{array}{c}\Sigma_b B(1/2^-) \\M_{th}=11090 {\rm MeV}\end{array}$} 
&$0.5$ &$1.6+0.00i$ &\multicolumn{1}{c}{$--$ } &\multicolumn{1}{c}{$--$ }     &\multicolumn{1}{c}{$--$ }  &$1.4+0.00i$     &$1.4+0.00i$  \\
&$1.0$ &$6.0+0.04i $ &\multicolumn{1}{c}{$--$ }  &\multicolumn{1}{c}{$--$ }  &\multicolumn{1}{c}{$--$ }  &$5.3+0.04i$      &$5.3+0.00i$ \\
&$1.5$ &$11.8+0.33i $ &\multicolumn{1}{c}{$--$ } &\multicolumn{1}{c}{$--$ }       &\multicolumn{1}{c}{$--$ }  &$10.2+0.25i $     &$10.4+0.00i$ \\
&$2.0$ &$17.9+1.60i$ &\multicolumn{1}{c}{$--$ }   &\multicolumn{1}{c}{$--$ }     &\multicolumn{1}{c}{$--$ }  &$15.4+1.17i$     &$16.1+0.00i$ \\
\bottomrule[2pt]
\end{tabular}

\end{table*}

In the first column, we list thresholds with certain spin parity, and the result of pole under the corresponding threshold with different $\alpha$ is given in the second and third columns with full coupled-channel $\Sigma^{(*)}_bB^{(*)}-\Lambda_bB^{(*)}$ interaction.  One can find, except a small width is acquired, the results are similar to those from the single-channel calculation. Hence, we take such channel as  production channel of this pole. All poles appear on threshold at about $\alpha=0.5$ and leave the threshold with the increasing of  $\alpha$. If we choose a binding energy about 10 MeV, the widths of most states are very small,  about 1 MeV or smaller. 

In the fourth to  eighth columns, we consider  two-channel result with the coupling between the production channel and  a channel below it.  The imaginary part reflects the strength of couplings between two channels.  Since  the pole is mainly from the production channel,  the effect of a channel on the pole can be also estimated from the two-channel resutls. Because the width with smaller $\alpha$ is very small, in the followings, we focus on the results at larger $\alpha$, 1.5 and 2.0.

Three states near $\Sigma^*_b B^*$ threshold, which is the highest threshold of channels considered in the current work,  can decay into five channels. Among these decay channels, the $\Lambda_bB^*$ channel has strongest couplings to this three states. For  the two states  near $\Sigma_b B^*$ threshold, there are four decay channels,  the $\Lambda_bB^*$ channel is much stronger than other channels for $3/2^-$. For $1/2$ state, both $\Sigma^*_bB$ and $\Lambda_bB^*$ channels couples strongly to the $\Sigma_b B^*$ channel.  Among three possible decay channels of  the state near $\Sigma^*_bB$ thresholds, only $\Lambda_bB^*$ channel provides large width, and  other channels only give  very small imaginary part of the position.  For the $\Sigma_bB(1/2^-)$ case, only $\Lambda_bB^{(*)}$ channels involves, among which, the $\Lambda_bB^*$ channel is still  dominant one.  Hence, for all seven states, the $\Lambda_bB^*$ channel is the most important one in all channels considered, which is consistent with the results of the hidden-charm pentaquarks in Refs.~\cite{Lin:2018kcc,Lin:2019qiv,He:2019rva}.

\section{Summary and discussion}\label{5}

In this work, the masses and widths of  hidden-bottom molecular pentaquarks are predicted from coupled-channel $\Sigma^{(*)}_bB^{(*)}-\Lambda_bB^{(*)}$ interaction in the qBSE approach with the help of effective Lagrangians with heavy quark and chiral symmetries. The results suggest that seven molecular states can be produced from the interactions. All states appear at $\alpha$ about 1, which corresponds to reasonable radii of the constituent hadrons, about 0.5~fm. 
 
Among the seven states, three of them are near the $\Sigma_b^*B^*$ threshold, and the masses of these three states are very close. The two sates with $1/2^-$ and $3/2^-$ are very weak compared with the state with $5/2^-$.  Hence, these three states should exhibit as one resonance   structure without partial wave analysis. Even with  partial wave analysis, the states with $1/2^-$ and $3/2^-$ are difficult to be distinguished from the one with $5/2^-$.  The two states near $\Sigma_bB^*$ are also mixing together but a partial-wave analysis will be helpful to distinguish them.  Hence, the results suggest that four resonance structures may be observed in experiment, though there exist seven molecular states from coupled-channel $\Sigma^{(*)}_bB^{(*)}-\Lambda_bB^{(*)}$ interaction.

Compared with hidden-charm $P_c$ states, the widths of  $P_b$ states are much smaller, about 1 MeV or smaller. And the calculation suggests that the  $\Lambda_bB^*$ channel has strong couplings to the molecular states, which is due to the strong couplings of vertex $\Lambda_b \Sigma_b^{(*)}\pi$. The small width have both advantage and disadvantage in experimental observation of such states. The small width makes the production possibility small, which needs high luminosity of  experimental facility. However, a small width also makes the peak of state stand out obviously from background in experiment. In Ref.~\cite{{Wang:2019zaw}}, we study the possibility to search for such states in pion and photon induced productions. The results suggest that with  small widths the measurement of the $P_b$ states is promising at the such as the COMPASS J-PARC, especially  the Electron Ion Collider (EicC) in China.

\vskip 10pt

\noindent {\bf Acknowledgement} This project is supported by the National Natural Science
Foundation of China with Grants No. 11675228.

%


\end{document}